\documentclass[12pt]{article}
\pdfoutput=1
\usepackage[round,longnamesfirst]{natbib}
\usepackage{csquotes,amsmath, amsthm, amssymb, amsbsy, mdwlist,ifthen}

\usepackage[bookmarksnumbered=true,backref=page,hyperfigures=true]{hyperref}

\hypersetup{
pdfauthor = {Pavel N. Krivitsky; Mark S. Handcock},
pdftitle = {A Separable Model for Dynamic Networks},
pdfkeywords = {Social networks; Longitudinal; Exponential random graph model; Markov chain Monte Carlo; Maximum likelihood estimation}
}

\def\y{{\boldsymbol{y}}}
\def\Y{{\boldsymbol{Y}}}
\def\x{{\boldsymbol{x}}}

\def\natpar{{\boldsymbol{\eta}}}
\def\curvpar{{\boldsymbol{\theta}}}
\def\genstat{{\boldsymbol{g}}}

\def\nactors{ n }
\def\actors{ N }

\def\setsub{\backslash}

\def\dysY{\mathbb{Y}}
\def\netsY{\mathcal{Y}}

\def\formsymb{{+}}
\def\disssymb{{-}}

\def\netsYF{\netsY^\formsymb}
\def\netsYD{\netsY^\disssymb}

\DeclareMathOperator{\Prob}{Pr}
\def\int{\lambda}
\def\Peg{\Prob_{\natpar,\genstat}}
\def\PegF{\Prob_{\natparF,\genstatF}}
\def\PegD{\Prob_{\natparD,\genstatD}}

\DeclareMathOperator{\Odds}{Odds}
\def\Oeg{\Odds_{\natpar,\genstat}}
\def\ceg{c_{\natpar,\genstat}}
\def\cegF{c_{\natparF,\genstatF}}
\def\cegD{c_{\natparD,\genstatD}}
\def\YF{\Y^\formsymb}
\def\YD{\Y^\disssymb}
\def\yF{\y^\formsymb}
\def\yD{\y^\disssymb}

\def\argmax{\arg\max}

\DeclareMathOperator{\ilogit}{logit^{-1}}
\def\RR{\mathbb{R}}
\def\NN{\mathbb{N}}
\def\pij{(i,j)}

\def\ijdysY{\pij\in\dysY}

\def\sij{_{i,j}}

\def\intij{\int\sij}

\def\intFij{\int^\formsymb\sij}
\def\intDij{\int^\disssymb\sij}
\def\integij{\int_{\natpar,\genstat,\pij}}

\def\yij{\y\sij}

\def\yFij{\y^\formsymb\sij}

\def\half{\frac{1}{2}}

\def\etheta{\natpar(\curvpar)}
\def\ethetaF{\natparF(\curvparF)}
\def\ethetaD{\natparD(\curvparD)}
\def\curvparF{\curvpar^\formsymb}
\def\curvparD{\curvpar^\disssymb}
\def\natparF{\natpar^\formsymb}
\def\natparD{\natpar^\disssymb}
\def\genstatF{\genstat^\formsymb}
\def\genstatD{\genstat^\disssymb}
\def\sigsym{{\tiny$\star$}}
\newcommand{\sig}[1]{\raisebox{0.5em}{\ifthenelse{\equal{#1}{0}}{\phantom{\sigsym\sigsym\sigsym}}{}\ifthenelse{\equal{#1}{1}}{\sigsym\phantom{\sigsym\sigsym}}{}\ifthenelse{\equal{#1}{2}}{\sigsym\sigsym\phantom{\sigsym}}{}\ifthenelse{\equal{#1}{3}}{\sigsym\sigsym\sigsym}{}}}

\newcommand{\ethetaS}[1]{\natpar(\curvpar^{#1})}

\newcommand{\myexp}[1]{\exp\left(#1\right)}

\newcommand{\I}[1]{1_{#1}}
\newcommand{\pkg}[1]{\textbf{#1}}
\newcommand{\proglang}[1]{\textsf{#1}}

\newcommand{\yat}[1]{\y^{t#1}}
\newcommand{\Yat}[1]{\Y^{t#1}}
\newcommand{\Yyat}[1]{\Yat{#1}=\yat{#1}}

\newcommand{\yatij}[1]{\y^{t#1}\sij}
\newcommand{\Yatij}[1]{\Y^{t#1}\sij}
\newcommand{\Yyatij}[1]{\Yatij{#1}=\yatij{#1}}

\providecommand{\abs}[1]{\left\lvert#1\right\rvert}

\theoremstyle{definition}
\newtheorem*{defn*}{Definition}

\title{A Separable Model for Dynamic Networks}
\author{Pavel N. Krivitsky\footnote{Research Associate,
Department of Statistics (E-mail:
\texttt{\href{mailto:krivitsky@stat.psu.edu}{krivitsky@stat.psu.edu}}), Penn State University, University Park, PA 16802}
\and Mark S. Handcock\footnote{Professor of
Statistics, Department of Statistics (E-mail:
\texttt{\href{mailto:handcock@stat.ucla.edu}{handcock@stat.ucla.edu}}), University of California, Los Angeles, CA
90095}}
\date{\today}
\begin{document}
\maketitle

\begin{abstract}
Models of dynamic networks --- networks that evolve over time ---
have manifold applications. 
We develop a discrete-time generative model
for social network evolution that inherits the richness
and flexibility of the class of exponential-family random graph models.
The model --- a Separable Temporal ERGM (STERGM) --- facilitates separable modeling of the tie duration
distributions and the structural dynamics of tie formation.
We develop likelihood-based inference for the model, and provide 
computational algorithms for maximum likelihood estimation.
We illustrate the interpretability of the model in analyzing 
a longitudinal network of friendship ties within a school.
\end{abstract}
\vspace*{.3in}

\section{Introduction}
Relational phenomena occur in many fields and are increasingly being
represented by networks.  There is a need for realistic and tractable
statistical models for these networks, especially when the phenomena
evolves over time.  For example, in epidemiology there is a need for
data-driven modeling of human sexual relationship networks for the
purpose of modeling and simulation of the spread of sexually
transmitted disease. As \citet{morris1997cps} show, spread of such
disease is affected not just by the momentary number of partnerships,
but their timing. To that end, the models used must have realistic
temporal structure as well as cross-sectional structure.

\citet{holland1977dms}, \citet{frank1991sac}, and others describe
continuous-time Markov models for evolution of social networks. (See
\citet{doreian1997esn} for a review.)  The most popular
parametrisation is the \emph{actor-oriented} model described by
\citet{snijders2005mln} and \citet{snijdersetal2010}, which can be
viewed in terms of actors making decisions to make and withdraw ties
to other actors. This model was then extended by
\citet{snijders2007mcn} to jointly model actors' network-related
choices (\enquote{selection}) and the effects of neighboring actors on
each other's attributes (\enquote{influence}).

Exponential-family random graph models (ERGMs) for social networks are
a natural way to represent dependencies in cross-sectional graphs and
dependencies between graphs over time, particularly in a discrete
context. \citet{robins2001rgm} first described this
approach. \citet{hanneke2007dtm} and \citet{hanneke2010dtm} also
define and describe a \emph{Temporal ERGM (TERGM)} (\enquote{Discrete
  Temporal ERGM} in the \citeyear{hanneke2007dtm} publication),
postulating an exponential family model for the transition probability
from a network at time $t$ to a network at time $t+1$.

Most of the attention in modeling of dynamic networks has focused on
fitting the model to a network series
\citep{snijders2001ses,hanneke2007dtm,hanneke2010dtm} or an
enumeration of instantaneous events between actors in the network
\citep{butts2008ref}. In the former case, the dyad census of the
network of interest is observed at multiple time points. In the latter
case, each event of interest and its exact time of occurrence is
observed.
  
A primary issue in modeling dynamic networks that has received limited
attention is that of attribution of prevalence. A snapshot of a
network at a single time point provides information about
\emph{prevalence} of the network properties of interest --- such as
the total number of ties --- as opposed to properties of a dynamic
network process that has produced it: \emph{incidence} --- the rate at
which new ties are formed --- and \emph{duration} --- how long they
tend to last once they do. Multiple snapshots over the same set of
actors (panel data) contain information about incidence and duration,
but, as we show below, the model parametrisations presently in use do
not allow convenient control over this attribution of prevalence.
 
In Section~\ref{sec:review}, we review discrete-time ERGM-based
network models, and in Section~\ref{sec:sep}, we extend these network
models to provide a more interpretable and convenient parametrisation
that separates incidence from duration. In Section~\ref{sec:cond-est},
we develop conditional maximum likelihood estimators (CMLE) based on
regularly-spaced network series data by extending the approach of
\citet{hunter2006ice}.  In Section~\ref{sec:example}, we illustrate
the methodology with application to a longitudinal network of
friendship ties within a school.  In Section~\ref{sec:Discussion}, we
consider some extensions that the model framework suggests and allows.

\section{Discrete-Time ERGM-Based Models for Network Evolution}\label{sec:review}
We first consider a discrete-time dynamic network model in which the
network at time $t$ is a single draw from an ERGM conditional on the
network at time $t-1$ (and possibly time $t-2$, etc.), extending the
Temporal ERGM (TERGM) of \citet{hanneke2007dtm} and
\citet{hanneke2010dtm}.  In this section we specify the model and
discuss its fundamental properties.

\subsection{Model Definition}\label{sec:review-dtergm}
Suppose that $\actors$ is the set of $\nactors=\abs{\actors}$ actors
of interest, labeled $1,\dotsc, \nactors$, and let
$\dysY\subseteq\actors\times\actors$ be the set of potential ties
among them --- with pairs $\ijdysY$ ordered for directed and unordered
for undirected networks --- and let $\netsY\subseteq 2^\dysY$ be the
set of possible networks of interest formed among these actors. For a
network realization $\y\in\netsY$, define $\yij$ to be an indicator of
a tie from actor $i$ to actor $j$, and further let $\y_{i,\cdot}$ be
the set of actors to whom $i$ has a tie, $\y_{\cdot,j}$ the set of
actors who have ties to $j$, and $\y_i$ the set of actors with
undirected ties with $i$.  Let $\Yat{} \in \netsY$ be a random
variable representing the state of the network at the discrete time
point $t$ and $\yat{}\in\netsY$ be its realization.
  
Following \citet{hunter2006ice}, let $\curvpar\in\RR^q$ be a vector of
$q$ model parameters, and let $\etheta:\RR^q\to\RR^p$ be a mapping
from $\curvpar$ to natural parameters $\natpar\in\RR^p$, with $q\le
p$. Let $\genstat:\netsY^2\to\RR^p$ be the sufficient statistic for
the transition from network $\yat{-1}$ at time $t-1$ to network
$\yat{}$ at time $t$. The one-step transition probability from
$\yat{-1}$ to $\yat{}$ is then defined to be
\begin{equation}
  \Peg(\Yyat{}|\Yat{-1}=\yat{-1};\curvpar)=\frac{\myexp{\etheta\cdot\genstat(\yat{}, \yat{-1})}}{\ceg(\curvpar,\yat{-1})},\ \yat{},\yat{-1}\in\netsY,\label{eq:dtergm}
\end{equation}
or, with a $k$-order Markov assumption, and letting $\genstat:\netsY^{k+1}\to\RR^p$, \begin{multline}
  \Peg(\Yyat{}|\Yyat{-1},\dotsc,\Yyat{-k};\curvpar)=\\
  \frac{\myexp{\etheta\cdot\genstat(\yat{}, \yat{-1},\dotsc,\yat{-k})}}{\ceg(\curvpar,\yat{-1},\dotsc,\yat{-k})},\ \yat{},\yat{-1},\dotsc,\yat{-k}\in\netsY,\label{eq:dtergm-k}
\end{multline}
and
\[\ceg(\curvpar,\yat{-1},\dotsc,\yat{-k})=\sum_{\y'\in\netsY}\myexp{\etheta \cdot\genstat(\y',\yat{-1},\dotsc,\yat{-k})},\] the normalizing constant.

TERGMs are a natural elaboration of the traditional ERGM
framework. They are essentially stepwise ERGM in time.  Note that the
definitions of \citet{robins2001rgm} and \citet{hanneke2007dtm} used
linear ERGMs only, where $\etheta\equiv\curvpar$ and $p\equiv q$.  To
simplify notation, from this point on we suppress reference to
$\natpar$ and $\genstat$.  

\def\Peg{\Prob}
\def\PegF{\Prob}
\def\PegD{\Prob}

\subsection{\label{sec:dtergm-interp}Model Specification and Interpretation}
The class of models specified by \eqref{eq:dtergm} is very broad and a
key component of model specification is the selection of $\genstat$.
Natural candidates are those developed for cross-sectional networks,
such as those enumerated by \citet{morris2008ser}.  However, the
choices in this dynamic situation are richer and can be any valid
network statistics evaluated on $\yat{}$ especially those that depend
on $\yat{-1}$.  \citet{hanneke2007dtm} focused on a choice of
$\genstat$ that had the property of \emph{conditional dyadic
  independence} --- that
\begin{equation}
  \Peg(\Yyat{}|\Yyat{-1};\curvpar)=\prod_{\ijdysY} \Peg(\Yyatij{}|\Yyat{-1};\curvpar),\label{eq:cond-dyad-ind}
\end{equation}
the distribution of $\Yat{}$ in which tie states are independent, but
only conditional on \emph{the whole of} $\Yat{-1}$.
  
However, caution must be used in interpreting their
parameters. Consider the simplest such statistic, the edge count:
\[\genstat(\yat{},\yat{-1})=\abs{\yat{}}.\] A higher coefficient on
$\genstat$ will, for any $\yat{-1}$, produce a $\Yat{}$ distribution
in which networks with more ties have higher probability. But, note
that this term would accomplish it in two ways simultaneously: it
would both increase the weight of those networks in which more ties
were formed on previously empty dyads and increase the weight of those
networks in which more extant ties were preserved (fewer
dissolved). That is, it would both increase the incidence and increase
the duration.
  
\citet{hanneke2007dtm} gave an example of a statistic that controls
the rate of evolution of the network: a measure of \emph{stability}.
This statistic counts the number of tie variables whose states did not
change between time steps, which is then divided by the maximum number
of ties an actor could have (a constant):
  \[\genstat(\yatij{},\yatij{-1})=\frac{1}{\nactors-1}\sum_{\ijdysY}
\left(\yatij{}\yatij{-1}+(1-\yatij{})(1-\yatij{-1})\right).\] A higher
coefficient on it will slow the evolution of the network down and a
lower coefficient will speed it up. From the point of view of
incidence and duration, however, it will do so in two ways: a higher
coefficient will result in networks that have fewer new ties formed
and fewer extant ties dissolved --- incidence will be decreased and
duration will be increased.
  
The two-sided nature of these effects tends to muddle parameter
interpretation, but a more substantial issue arises if selective
mixing statistics, like those described by \citet{koehly2004efm}, are
used. Consider a concrete example, with actors partitioned into $K$
known groups, with $\mathbb{K}\subseteq \{1,\dotsc,K\}^2$ being the
set of pairs of groups between whose actors there may be ties. (For
example, in a directed network, $\mathbb{K}=\{1,\dotsc,K\}^2$.)  Let
$P_k$ be the set of actors who belong to group $k$ and $P(i)$ be the
partition to which actor $i$ belongs. The model with transition
probability
  \begin{multline} \Peg(\Yyat{}|\Yyat{-1};\curvpar)\propto\\ \exp
\left(\curvpar_0 \sum_{\ijdysY}
\left(\yatij{}\yatij{-1}+(1-\yatij{})(1-\yatij{-1})\right)+\sum_{(k_1,k_2)\in
\mathbb{K}}\curvpar_{k_1,k_2}|\y_{P_{k_1},P_{k_2}}|\right), \label{eq:churning-model}
  \end{multline} models stability, controlled by $\curvpar_0$, and
mixing among the groups, controlled by $\curvpar_{k_1,k_2}$. (Here,
$|\y_{P_{k_1},P_{k_2}}|$ is defined as the number of ties from actors
in group $k_1$ to actors in group $k_2$ for directed networks, and
ties between actors in those groups for undirected networks.)
  
  Given $\yat{-1}$, the probability that a given non-tied directed
pair $\pij $ will gain a tie in a given time step is
  \[\Peg(\Yatij{}=1|\Yatij{-1}=0;\curvpar)=\ilogit(-\curvpar_0+\curvpar_{P(i),P(j)}),\]
and the probability that an extant tie $\pij $ will be removed is
  \[\Peg(\Yatij{}=0|\Yatij{-1}=1;\curvpar)=\ilogit(-\curvpar_0-\curvpar_{P(i),P(j)}),\]
the latter leading to a duration distribution which is geometric with
support $\NN$ and expected value \citep*[pp. 621--622]{casella2002si}
  \[\left(\ilogit(-\curvpar_0-\curvpar_{P(i),P(j)})\right)^{-1}=1+\myexp{\curvpar_0+\curvpar_{P(i),P(j)}}.\]
Thus, a higher value of coefficient $\curvpar_{k_1,k_2}$
simultaneously increases the incidence of ties between actors in group
$k_1$ and actors in group $k_2$ and their duration.
  
This coupling between the incidence of ties and their duration not
only makes such terms problematic to interpret, but has a direct
impact on modeling.  Consider a sexual partnership network, possessing
strong ethnic homophily, with ties within each ethnic category being
more prevalent (relative to the potential number of ties) than ties
between ethnic categories. (A real-world illustration of this effect
was given by \citet{krivitsky2011ans}.) This structure could be a
consequence of the within-ethnic ties being formed more frequently
than between-ethnic ties, of the within-ethnic ties lasting, on
average, longer than between-ethnic ties, or some combination of the
two.  With cross-sectional data alone, it is impossible to tell these
apart and a model like \eqref{eq:churning-model} implies a dynamic
process in which cross-ethnic ties toggle unnaturally frequently, or
\enquote{churn}.  We refer to a model with this dynamic pathology as a
\enquote{churning model} as this stochastic property is unlikely to be
seen in real phenomena.  Churning is related to the degeneracy
properties of ERGM \citep{hannas03}.

\section{Separable Parametrisation}\label{sec:sep}
We now motivate and describe the concept of \emph{separability} of
formation and dissolution in a dynamic network model, and describe the
\emph{Separable Temporal ERGM} (STERGM).
\subsection{Motivation}\label{sec:sep-motive}
Intuitively, those social processes and factors that result in ties
being formed are not the same as those that result in ties being
dissolved. For example, in the above-mentioned sexual partnership
network, the relative lack of cross-ethnic ties may be a result of
racial segregation, language and cultural barriers, racism, and
population-level differences in socioeconomic status, all of which
have a strong effect on the chances of a relationship forming.  Once
an interracial relationship has been formed, however, either because
these factors either did not apply in that case or were overcome, the
duration of such a relationship would likely not be substantially
lower. Even if it were lower, the differences in the probability of
such a relationship ending during a particular time interval would
not, in general, be a perfect reflection the differences in the
probability of it forming during such a time interval.

Furthermore, it is often the case in practice that information about
cross-sectional properties of a network (i.e. \emph{prevalence}) has a
different source from that of the information about its longitudinal
properties (i.e. \emph{duration}), and it may be useful to be able to
consider them separately \citep{krivitsky2008smd,krivitsky2009sms}.

Thus, it is useful for the parametrisation of a model to allow
separate control over incidence and duration of ties and separate
interpretation, at least over the short run. (For any nontrivial
process, formation and dissolution would likely interact with each
other in the long run.)

\subsection{Model Specification}
In this section, we introduce a class of discrete-time models for
network evolution, which assumes that these processes are separable
from each other within a time-step. We consider a sub-class of models
based on the ERGM family, which inherits the interpretability and
flexibility of those processes.

\subsubsection{\label{sec:gensepmod}General Separable Models}
We represent networks as sets of ties, so given $\y,\y' \in \netsY$,
the network $\y\cup \y'$ has the tie $\pij$ if, and only if, $\pij$
exists in $\y$ or $\y'$ or both; the network $\y\cap \y'$ has $\pij$
if, and only if, $\pij$ exists in both $\y$ and $\y'$; and the network
$\y\setsub \y'$ has tie $\pij$ if, and only if, $\pij$ exists in $\y$
but not in $\y'$. The relation $\y \supseteq \y'$ holds, if, and only
if, $\y$ has all of the ties that $\y'$ does (and, possibly, other
ties as well), and conversely for $\y \subseteq \y'$.

Consider the evolution of a random network at time $t-1$ to time $t$,
and define two intermediate networks, the \emph{formation network}
$\YF$, consisting of the initial network $\Yat{-1}$ with ties formed
during the time step added and the \emph{dissolution network} $\YD$,
consisting of the initial network $\Yat{-1}$ with ties dissolved
during the time step removed (with $\yF$ and $\yD$ being their
respective realized counterparts). Then, given $\yat{-1}$, $\yF$, and
$\yD$, the network $\yat{}$ may be evaluated via a set operation, as
\begin{equation}
  \yat{}=    \yF\setsub(\yat{-1}\setsub\yD)=\yD\cup(\yF\setsub\yat{-1}).\label{eq:fdeval}
\end{equation}
Since it is the networks $\yat{-1}$ and $\yat{}$ that are actually
observed, $\yF$ and $\yD$ may be regarded as latent variables, but it
is possible to recover them given $\yat{-1}$ and $\yat{}$, because a
tie variable can only be in one of four states given in
Table~\ref{tab:Sim-dyad-states}. Each possibility has a unique
combination of tie variable states in $\yat{-1}$ and $\yat{}$, so
observing the network at the beginning and the end allows the two
intermediate states to be determined as $\yF=\yat{-1}\cup \yat{}$ and
$\yD=\yat{-1}\cap \yat{}$.
\begin{table}
  \caption{\label{tab:Sim-dyad-states}Possible transitions of a single tie variable}
  \centering
  \begin{tabular}{ccccc}
    \hline
    $\yatij{-1}$&$\rightarrow$&$(\yF\sij,\yD\sij)$&$\rightarrow$&$\yatij{}$\\
    \hline
0&$\rightarrow$&$(0,0)$&$\rightarrow$&0\\
0&$\rightarrow$&$(1,0)$&$\rightarrow$&1\\
1&$\rightarrow$&$(1,0)$&$\rightarrow$&0\\
1&$\rightarrow$&$(1,1)$&$\rightarrow$&1\\
\hline
  \end{tabular}
\end{table}

If $\YF$ is conditionally independent of $\YD$ given $\Yat{-1}$ then
\begin{multline}
  \Prob(\Yyat{}|\Yyat{-1};\curvpar)=\\
  \PegF(\YF=\yF|\Yyat{-1};\curvpar )\times\PegD(\YD=\yD|\Yyat{-1};\curvpar )\label{eq:condind}
\end{multline}
We refer to the two factors on the RHS as the \emph{formation model}
and the \emph{dissolution model}, respectively.  Suppose that we can
express $\curvpar=(\curvparF ,\curvparD )$ where the formation model
is parametrised by $\curvparF$ and the dissolution model by
$\curvparD$.
\begin{defn*}
  We say that a dynamic model is \emph{separable}
  if $\YF$ is conditionally independent of $\YD$ given $\Yat{-1}$ and
  the parameter space of $\curvpar$ is the product of the individual
  parameter spaces of $\curvparF$ and $\curvparD$.
\end{defn*}

We refer to such a model as separable because it represents an
assumption that during a given discrete time step, the process by
which the ties form does not interact with the process by which they
dissolve: both are separated (in the conditional independence sense)
from each other conditional on the state of the network at the
beginning of the time step.  

\subsubsection{Generative Mechanism}
Let some $\netsYF(\yat{-1})\subseteq \{\y\in
2^\dysY:\y\supseteq\yat{-1}\}$ be the sample space, under the model,
of formation networks, starting from $\yat{-1}$; and let some
$\netsYD(\yat{-1})\subseteq \{\y\in 2^\dysY:\y\subseteq\yat{-1}\}$ be
the sample space of dissolution networks.  The model postulates the
following process for evolution of a random network at time $t-1$ to a
random network at time $t$:
\begin{enumerate*}
\item Draw an intermediate network $\yF$ from the distribution
  \[\PegF(\YF = \yF|\Yyat{-1};\curvparF ),\ \yF\in\netsYF(\yat{-1}).\]
\item Draw an intermediate network $\yD$ from the distribution 
  \[\PegF(\YD = \yD |\Yyat{-1};\curvparD ),\ \yD\in\netsYD(\yat{-1}).\]
\item Apply formations and dissolutions to $\yat{-1}$ to produce
  $\yat{}$ by evaluating (\ref{eq:fdeval}).
\end{enumerate*}
Note that, as specified, this model is first order Markov, but
$\Yat{}$ can be further conditioned on $\Yat{-2}$, $\Yat{-3}$, etc, to
produce higher order versions.  We do not develop these models here.

\subsubsection{Separable Temporal ERGM (STERGM)}\label{sec:sep-ergm}
A natural family of models for the components of the separable model
is the ERGMs considered in Section 2.1. We focus on this rich class of
models in the remainder of the paper.  Specifically, we model:
\begin{equation*}\PegF(\YF = \yF|\Yyat{-1};\curvparF )=\frac{\myexp{\ethetaF\cdot \genstatF(\yF,\yat{-1})}}{\cegF(\curvparF ,\yat{-1})},\ \yF\in\netsYF(\yat{-1})\end{equation*}
and
\begin{equation*}\PegD(\YD = \yD|\Yyat{-1};\curvparD )=\frac{\myexp{\ethetaD\cdot \genstatD(\yD,\yat{-1})}}{\cegD(\curvparD ,\yat{-1})},\ \yD\in\netsYD(\yat{-1}),\end{equation*} 
with their normalizing constants $\cegF(\curvparF ,\yat{-1})$ and
$\cegD(\curvparD ,\yat{-1})$ summing over $\netsYF(\yat{-1})$ and
$\netsYD(\yat{-1})$, respectively.

We now derive the probability of transitioning from a given network at
time $t-1$, $\yat{-1}$ to a given network at time $t$, $\yat{}$.
Based on \eqref{eq:condind}, we have
\[\Peg(\Yyat{}|\Yyat{-1};\curvpar)=\frac{\myexp{\etheta\cdot\genstat(\yat{}, \yat{-1})}}{\cegF(\curvparF ,\yat{-1})\cegD(\curvparD ,\yat{-1})}.\]
where $\natpar=(\natparF,\natparD)$ and
$\genstat(\yat{},\yat{-1})=(\genstatF(\yat{-1}\cup
\yat{},\yat{-1}),\genstatD(\yat{-1}\cap \yat{},\yat{-1}))$.  As
$\Peg(\Yyat{}|\Yyat{-1};\curvpar)$ is, by construction, a valid
probability mass function, \[\cegF(\curvparF ,\yat{-1})\cegD(\curvparD
,\yat{-1})=\ceg(\curvpar,\yat{-1}),\] where
\[\ceg(\curvpar,\yat{-1})=\sum_{\y'\in\netsY}\myexp{\etheta \cdot\genstat(\y',\yat{-1})}.\]
This is the same form as~\eqref{eq:dtergm}.  Thus, the STERGM class is
a subclass of a first-order Markov TERGM of \citet{hanneke2007dtm},
described in Section~\ref{sec:review-dtergm}: any transition process
that can be expressed with $\genstatF$, $\genstatD$, $\natparF$ and
$\natparD$ can be reproduced by a model in the TERGM class.  However,
the essential issue is the specification of models within these
classes, and the value of the STERGM class is that it focuses
specification on a viable and fecund region in the very broad class.
In the parametrisation in terms of formation and dissolution, some
flexibility is lost --- the ability to have the formation and
dissolution processes interact within a given time step. What is
gained is ease of specification, tractability of the model, and
substantial improvement in interpretability.

\subsection{Interpretation} \label{sec:sep-interp}

In contrast to statistics like \emph{stability} in
Section~\ref{sec:dtergm-interp}, the STERGM's sufficient statistics
and parameters have an implicit direction: they affect directly either
incidence or duration, but not both, and even statistics that do not
explicitly incorporate the previous time step's network $\yat{-1}$,
incorporate it via the constraint of the phase in which they are
used. This allows familiar cross-sectional ERGM sufficient statistics
to be used, with their parameters acquiring intuitive interpretations
in terms of the network evolution process. We call these inherited
terms, for which $\genstatF_k(\yF,\yat{-1})\equiv\genstatF_k(\yF)$
and/or $\genstatD_k(\yD,\yat{-1})\equiv\genstatD_k(\yD)$, with no
further dependence on $\yat{-1}$, \emph{implicitly dynamic}.

Such terms (and their corresponding coefficients) often have
straightforward general interpretations for formation and dissolution
phases. In particular, consider an implicitly dynamic statistic that
counts the number of instances of a particular feature found in the
network $\yF$ or $\yD$. Examples of features that might be counted
include a tie, an actor with exactly $d$ neighbors, or a tie between
an actor in a set $P_{k_1}$ and an actor in the set $P_{k_2}$.  

\subsubsection{Formation}
A positive $\curvparF_k$ corresponding to a particular $\genstatF_k$
increases the probability of those $\yF$ which have more instances of
the feature counted by $\genstatF_k$ --- greater values of
$\genstatF_k(\yF)$. This affects the network process in two ways: the
probability of forming those ties that create new instances of the
feature counted by $\genstatF_k$ is increased and the probability of
forming those ties that \enquote{disrupt} those instances would be
reduced.

Conversely, negative $\curvparF _k$ would result in higher
probabilities for those networks with fewer instances of the feature
counted by $\genstatF_k$, reducing the probability of forming ties to
create more instances of the feature counted and increasing the
probability of forming ties to \enquote{disrupt} the feature.
  
Notably, $\genstatF_k$ counts features in the network
$\yF=\yat{}\cup\yat{-1}$, rather than in the ultimately observed
network $\yat{}$. This means that for some features, particularly
those with dyadic dependence, the dissolution process may influence
the feature so that it is present in $\yF$ but not in $\yat{-1}$ or
$\yat{}$. How frequently this occurs depends on the specific model and
the rate of evolution of the network process: if a network process is
such that the network changes little (in both formation and
dissolution) during each time step, such interference is unlikely.
\subsubsection{Dissolution}
As in the formation phase, a positive $\curvparD_k$ corresponding to a
particular $\genstatD_k$ increases the probability of those $\yD$
which have more instances of the feature counted by $\genstatD_k$,
thus tending to preserve more instances of that feature (or dissolving
ties to create more instances, as may be the case with
dyadic-dependent terms), while a negative $\curvparD_k$ will increase
the probability of networks with fewer instances of the feature in
question, effectively causing the dissolution process to target those
features, and also refrain from dissolving ties whose dissolution
would create those features. It is important to note that the
dissolution phase ERGM determines which ties are \emph{preserved}
during the time step, and the parameters should be interpreted
accordingly.

Again, it is $\yD=\yat{}\cap\yat{-1}$ on which statistics are
evaluated, so the formation process can interact with the dissolution
process as well.

These principles mean that many of the vast array of network
statistics developed for ERGMs \citep[for example]{morris2008ser} can
be readily adapted to STERGM modeling, retaining much of their
interpretation.  In the Appendix, we develop and give interpretations
to the fundamental edge count, selective mixing by actor attribute,
and degree distribution terms.
\subsubsection{Explicitly Dynamic Terms}
At the same time, some effects on formation and dissolution may depend
on specific features of $\yat{-1}$. For instance, consider a social
process in which an actor having multiple partners (e.g.,
\enquote{two-timing}) is actively punished, so having more than one
partner in $\yat{-1}$ increases the hazard of losing all of one's
partners in $\yat{}$. (Such an effect may be salient in a sexual
partnership network.) This dissolution effect cannot be modeled by
implicitly dynamic terms, because it cannot be reduced to merely
increasing or reducing the tendency of $\YD$ to have particular
features. For example, a positive coefficient on a statistic counting
the number of actors with no partners (isolates) would increase the
weight of those $\yD$ that have more isolates, affecting the
dissolution of the sole tie of an actor with only one partner just as
much as it would affect the dissolution of ties of an actor with more
than one partner.

On the other hand, an \emph{explicitly dynamic} model term that counts
the number of actors with no partners in $\yD$ \emph{only among those
  actors who had two or more partners in $\yat{-1}$} would, with a
positive coefficient, increase the probability of a transition
directly from having two partners to having none. Beyond that, its
interpretation would be no different than that of an implicitly
dynamic dissolution term.
\subsection{Continuous-Time Markov Models}
Although the focus of this paper is on discrete-time models for
network evolution, the separability paradigm can be applied to
continuous-time network evolution models such as those of
\citet{holland1977dms}. There, network evolution is modeled as a
continuous-time Markov process such that the intensity of transition
between two networks that differ by more than one dyad is $0$, while
the evolution of the network is controlled by
$\int(\yat{};\curvpar):\netsY\to \RR_+^\dysY$, with each
$\intij(\yat{};\curvpar)$ being the intensity associated with toggling
each dyad $\pij$.

In that scenario, separation of formation and dissolution is realized
by formulating $\curvpar=(\curvparF ,\curvparD )$ and
\[\intij(\yat{};\curvparF ,\curvparD)=
\begin{cases}
  \intFij(\yat{};\curvparF) & \text{if $\yatij{}=0$} \\
  \intDij(\yat{};\curvparD) & \text{if $\yatij{}=1$}
\end{cases},
\]
with $\intFij(\yat{};\curvparF)$ and $\intDij(\yat{};\curvparD)$ being
formation- and dissolution-specific intensities. Indeed,
\citet{holland1977dms} use a formulation of this general
sort. Notably, unlike the discrete-time process, this separation
requires only separation of parameters and no additional independence
assumptions. This is because under the Markov assumption and with no
chance of more than one dyad toggling coincidentally at a specific
time, dyads effectively evolve independently in a sufficiently small
interval (i.e., $[t,t+h]$, $h\to 0$), and dyadic independence in
network evolution \emph{a fortiori} implies separability between which
ties form and which ties dissolve.

An exponential-family form for $\intij$,
\[\integij(\yat{};\curvpar)=
\begin{cases}
  \myexp{\ethetaF\cdot \genstatF(\yat{}\cup\{\pij\},\yat{})} & \text{if $\yatij{}=0$} \\
  \myexp{\ethetaD\cdot \genstatD(\yat{}\setsub\{\pij\},\yat{})} & \text{if $\yatij{}=1$}
\end{cases},
\]
may be viewed as the limiting case of the discrete-time STERGM, in
which the amount of time represented by each time step shrinks to
zero.

\section{Likelihood-Based Inference for TERGMs}\label{sec:cond-est}
In this section, we consider inference based on observing a series of
$T+1$ networks, $\y^0,\dotsc,\y^T$.  \citet{hanneke2007dtm} proposed
to fit TERGMs by finding the conditional MLE under an order $k$ Markov
assumption,
\begin{equation}
  \hat{\curvpar}=\argmax_\curvpar \prod_{t=k}^T\Peg(\Yyat{}|\Yyat{-k},\dotsc,\Yyat{-1};\curvpar),\label{eq:dtergm-mle-def}
\end{equation}
computing a Method-of-Moments estimator (equivalent in their case to
the MLE) with a simulated Newton-Raphson zero-finding algorithm.  We
extend the work of \citet{hunter2006ice} and \citet{geyer1992cmc} to
compute the conditional MLE for curved exponential-family transition
models (that is, cases where $\etheta\ne\curvpar$).

For simplicity, we consider models with first-order Markov
dependence. There is no loss of generality, since as long as the order
of Markov dependence $k$ is finite, we can define the depended-upon
network $\yat{-1}$ to implicitly \enquote{store} whatever information
about $\yat{-1},\dotsc,\yat{-k+1}$ is needed to compute the transition
probability.

The conditional MLE \eqref{eq:dtergm-mle-def} can then be obtained by
maximizing the log-likelihood
\[l(\curvpar)=\etheta\cdot \left(\sum_{t=1}^T\genstat(\yat{}, \yat{-1})\right) - \log \left(\prod_{t=1}^T\ceg(\curvpar,\yat{-1})\right).\]
For any two values of the model parameter $\curvpar^0$ and $\curvpar$, the log-likelihood-ratio is
\[l(\curvpar)-l(\curvpar^0)=\left(\etheta-\ethetaS{0}\right)\cdot
\left(\sum_{t=1}^T\genstat(\yat{}, \yat{-1})\right) - \log
\left(\prod_{t=1}^T\frac{\ceg(\curvpar,\yat{-1})}{\ceg(\curvpar^0,\yat{-1})}\right).\]
The main difficulty is in evaluating the ratio of the normalizing
constants.  These conditional normalizing constants depend on networks
at times $0,\dotsc,T-1$. However, these ratios can still be expressed
as
\begin{multline}
  \prod_{t=1}^T\frac{\ceg(\curvpar,\yat{-1})}{\ceg(\curvpar^0,\yat{-1})}=\prod_{t=1}^T\sum_{\y\in\netsY}\myexp{\left(\etheta-\ethetaS{0}\right)\cdot\genstat(\y, \yat{-1})}
  \times\frac{\myexp{\ethetaS{0}\cdot\genstat(\y, \yat{-1})}}{\ceg(\curvpar^0,\yat{-1})}\\
  =\prod_{t=1}^T\sum_{\y\in\netsY}\myexp{\left(\etheta-\ethetaS{0}\right)\cdot\genstat(\y, \yat{-1})}
  \times\Peg(\Yat{}=\y|\Yyat{-1};\curvpar^0).\label{eq:dtergm-mle-exp}
\end{multline}
The expression \eqref{eq:dtergm-mle-exp} is a product of expectations
over the conditional distribution under the model of $\Yat{}$ given
$\Yat{-1}$ at $\curvpar^0$, each of which can be estimated by
simulation, allowing the algorithm of \citet{hunter2006ice} to be
applied to fit a TERGM to network series data.

These results also make it possible to assess the goodness-of-fit of a
model via an analyses of deviance.  Specifically, we can compute the
change in log-likelihood from the null model $(\etheta=0)$ to the
conditional MLE. To do this, we extended the bridge sampler of
\citet{hunter2006ice} to this setting. 

\section{Application to the Dynamics of Friendship}\label{sec:example}
As an application of this model, we consider the friendship
relations among 26 students during their first year at a Dutch secondary school \citep{knecht2008}.
The friendship nominations were assessed at four time
points at intervals of three months starting at the beginning of their
secondary schooling. The friendship
data are directed and were assessed by asking students to indicate 
classmates whom they considered good
friends. 
There were 17 girls and 9 boys in the class. The data included covariates
collected on each student. Here, we consider the sex of the student, as it is a
primary determinant of the friendship ties. We also consider a dyadic covariate
indicating if each pair of students had gone to the same primary school.
These data were used to illustrate the actor-oriented
approach to modeling by \citet{snijdersetal2010} (whom we follow). 
That paper should be consulted
for details of the data set and an alternative analysis.

Some of the data at time points two through four were missing due to
student absence when the survey was taken.  These were accommodated
using the approach of \citet{hangile10} under the assumption that the
unobserved data pattern was amenable to the model.  One student left
the class after time point 1.  This could have been accommodated in a
number of ways, depending on the assumptions one is willing to make.
Here we considered the networks with this student omitted both as a
nominator and nominee of friendships.  As \citet{snijdersetal2010}
note, each student was allowed to nominate at most 12 classmates at
each time point. In general, inference needs to incorporate features
of sampling design such as this one.  We discuss how in
Section~\ref{sec:Discussion}. However, its effect here is
negligible: in the ($4\times 25=$) 100 student reports, only 3
nominated the maximum number.

Our objective is to explain the observed structural patterns of change in the
network over the course of the year. We build a model including both exogenous
and endogenous structural effects, following the same approach and motivations 
as \citet{snijdersetal2010}. 
For the formation component we include terms for the propensity of 
students to choose friends of the same or opposite sex (i.e., overall
propensities to nominate friends that are homophilous on sex or not). 
We include a term to
measure the propensity of friendships to be reciprocal. We include information
on the primary school co-attendance via a count of the number of times students
nominate other students with whom they went to primary school.  To capture any
overall propensity of students to nominate other students who are popular we
include an overall outdegree popularity effect \citep[equation (12)]{snijdersetal2010}. To model transitivity effects we include two terms. The first
is aggregate transitive ties that aims to capture a tendency toward 
transitive closure consistent with local hierarchy.
The second is an aggregate cyclical ties term to capture  anti-hierarchical
closure.
The terms in the model are structurally largely consistent with
the terms chosen in \citet{snijdersetal2010}.
A similar model was considered for the dissolution process.
Specifics of these terms are given in the Appendix.

We fit the model using the conditional MLE procedure of
Section~\ref{sec:cond-est}. Computationally this is implemented using
a variant of the MCMC approach of \citet{hunter2006ice}.  To monitor
the statistical properties of the MCMC algorithm we use the procedures
by \citet{hunter2008gfs}.  

Table~\ref{TableCMLEFN} reports
the estimates for the model assuming homogeneity of parameters over
time. The
outdegree popularity effect had a correlation of 0.995 with the edges
effect and was omitted from the model.
  \begin{table}
    \caption{MLE parameter estimates for the longitudinal friendship network
        \label{TableCMLEFN}}
    \centering
      \begin{tabular}{lrc@{\hspace*{1ex}}r}
        & \multicolumn{1}{c}{Formation\sig{0}} & &
        \multicolumn{1}{c}{Dissolution\sig{0}} \\
        Parameter     &   \multicolumn{1}{c}{est. (s.e.)\sig{0}} & &\multicolumn{1}{c}{est. (s.e.)\sig{0}} \\
        \hline
        Edges                       & $-3.336\ (0.320)$\sig{3}&& $-1.132\ (0.448)$\sig{1} \\
        Homophily (girls)           &  $0.480\ (0.269)$\sig{0}&&  $0.122\ (0.394)$\sig{0} \\
        Homophily (boys)            &  $0.973\ (0.355)$\sig{2}&&  $1.168\ (0.523)$\sig{1} \\
        F$\rightarrow$M heterophily & $-0.358\ (0.330)$\sig{0}&& $-0.577\ (0.609)$\sig{0} \\
        Primary school              &  $0.650\ (0.248)$\sig{2}&&  $0.451\ (0.291)$\sig{0} \\
        Reciprocity                 &  $1.384\ (0.280)$\sig{3}&&  $2.682\ (0.523)$\sig{3} \\
        Transitive ties             &  $0.886\ (0.247)$\sig{3}&&  $1.121\ (0.264)$\sig{3} \\
        Cyclical ties               & $-0.389\ (0.133)$\sig{2}&& $-1.016\ (0.231)$\sig{3} \\
        \hline
\\
\multicolumn{4}{l}{Significance levels: 0.05\sig{1}, 0.01\sig{2}, 0.001\sig{3}}
      \end{tabular}
  \end{table}

As for the standard ERGM, the individual $\theta$ coefficients can be interpreted as conditional log-odds ratios.
There is also a relative risk interpretation that is
often simpler.  For example, the exponential of the primary school coefficient is the relative risk of
formation or preservation (depending on the phase) of friendship between two students from the same primary school
compared to two students from different primary schools with the same values of the
other covariates and structural effects. The probabilities involved are
conditional on these other covariates and structural effects. The
interpretation for non-binary and multiple covariates
is similar: $\exp(\theta\Delta)$ is the relative risk of
friendship between two students compared to two students
with vector of covariates differing by $\Delta$ (and with
the same values of the other structural effects).

The standard errors of Table \ref{TableCMLEFN} 
are obtained from the information matrix in the likelihood
evaluated at the MLE to which we have added the 
(small) MCMC standard error obtained using the procedure given by
\citet{hunter2008epf}.

The networks at the earlier time points are strongly sexually
segregated, and we see strong homophily by sex in the formation of
ties. This effect is mildly stronger for boys than for girls. 
We do not see an overall disinclination for girls to nominate 
boys (relative to other combinations).
In other words, the boys are about as likely to form 
friendships as the girls.
As expected, we see a high degree of reciprocity in the formation of ties.
There is a strong transitive closure effect, with a positive coefficient on
transitive tie formation and a negative coefficient on cyclical tie formation.
This suggests a strong hierarchical tendency in the formation of ties.
We see that
students who attended the same primary school are much more likely to form
ties.

\begin{table}
  \caption{Analysis of deviance for the longitudinal friendship network, comparing time-homogeneous (hom.) and time-heterogeneous (het.) parametrisations \label{TableCMLEAoD}}
  \centering
    \begin{tabular}{lrrrcrrr}
      & \multicolumn{3}{c}{Formation}                          &  &  \multicolumn{3}{c}{Dissolution} \\
      \cline{2-4} \cline{6-8}
      & \multicolumn{1}{c}{residual} & \multicolumn{1}{c}{explained}\sig{0}&& & \multicolumn{1}{c}{residual} & \multicolumn{1}{c}{explained}\sig{0} \\
      Model           & \multicolumn{1}{c}{dev. (d.f.)}    & \multicolumn{1}{c}{dev. (d.f.)}\sig{0}& AIC  &&   \multicolumn{1}{c}{dev. (d.f.)}  & \multicolumn{1}{c}{dev. (d.f.)}\sig{0}& AIC \\
      \hline
      Null            & 1838 (1326)    &                      &1838  &&   459 (331)    &                      & 459 \\
      Edges (hom.)    &  924 (1325)    & 915 (\phantom{0}1)\sig{3}   & 926  &&   431 (330)    & 28 (\phantom{0}1)\sig{3}    & 433 \\
      Full (hom.)     &  819 (1318)    & 104 (\phantom{0}7)\sig{3}   & 835  &&   350 (323)    & 82 (\phantom{0}7)\sig{3}    & 366 \\
Full (hom. \\ 
~\hfill except edges) &  818 (1316)    &   2 (\phantom{0}2)\sig{0}   & 838  &&   344 (321)    & 6  (\phantom{0}2)\sig{0}    & 364 \\
      Full (het.)     &  795 (1302)    &  23 (14)\sig{0}             & 843  &&   314 (307)    &30 (14)\sig{2}               & 362 \\
      \hline
\\
\multicolumn{4}{l}{Significance levels: 0.05\sig{1}, 0.01\sig{2}, 0.001\sig{3}}
    \end{tabular}
\end{table}

These structural terms have less influence on the 
dissolution of ties.  
There is some modest evidence that boy to boy ties are
less likely to dissolve than other mixtures of sexes. 
(Recall that parameters represent a measure of persistence, so that negative
parameters are associated with shorter durations).
As expected, we see the dissolution of ties is strongly 
retarded by the presence of a reciprocal tie.
As in the formation process, there is a strong transitive closure effect
suggesting a strong hierarchical tendency in the dissolution of ties.
Once a hierarchical triad is formed it will tend to endure longer.
Students who attended the same primary school are not significantly more
likely to have persistent ties.

As the data measure a social process that is developing in time, we
do not need to assume that the process is in temporal equilibrium; thus we
could estimate separate parameters for the change between each pair of
successive time points. One such model specifies different overall rates
of tie formation or dissolution at each time point but retains homogeneous parameters for the
other terms. Another allows all the parameters to vary at each time point.

Table~\ref{TableCMLEAoD} gives the analysis
of deviance for formation and dissolution models nested above and below
those in Table~\ref{TableCMLEFN}.
For the formation process we see the full time-homogeneous model in 
 Table~\ref{TableCMLEFN} significantly improves on the null and 
Erd\H{o}s-R\'{e}nyi model (Edges (hom.)). Specifying different overall rates
of tie formation at each time point does not significantly improve the fit, nor
does a full time-heterogeneous model with different structural parameters at
each time point.
For the dissolution process,
we again see the full time-homogeneous model
significantly improves on the null and 
Erd\H{o}s-R\'{e}nyi model. However, there is some evidence that
specifying time-heterogeneous versions improve the fit.
An inspection of the time-heterogeneous models indicates that most of the
improvement is due to the increase in 
hierarchical tendency over time. Initially this transitive closure does not
retard tie dissolution, but it does over time.

\section{Discussion}\label{sec:Discussion}
This paper introduces a statistical model for networks
that evolve over time. It builds on the
foundations of exponential-family random graph models for
cross-sectional networks and inherits the flexibility and
interpretability of these models. In addition, it leverages the
inferential and computational tools that have been developed for
ERGMs over the last two decades.  

As we showed in Section~\ref{sec:review}, parameters used in models
currently in use directly affect both the incidence of ties (at a
given time point) and the duration of ties (over time).  STERGMs have
one set of parameters control formation of new ties and another
control dissolution (or non-dissolution) of extant ties. Such a
separable parametrisation controls the attribution of incidence and
duration and greatly improves the interpretability of the model
parameters, all without sacrificing the ability to explicitly
incorporate effects of specific features of past networks, if needed.

It is important to emphasize that STERGMs jointly model the
formation and dissolution of ties.  While the two processes are
modeled as conditionally independent within a time step, they are
modeled as dependent over time. More importantly, they allow the
structure of the incidence to be identified in the presence of
the durational structure.

In addition, the model has computational advantages.  The
likelihood function can be decomposed and the components computed
relatively easily.  All computations in this paper were completed
using the \pkg{ergm} \citep{hunter2008epf,handcock2011epf} package from the \pkg{statnet}
\citep{handcock2008sst} suite of libraries for social network
analysis in \proglang{R} \citep{rdevelopmentcoreteam2009rla}.

The model is directly applicable to both directed and undirected
networks.  It can be easily tuned to applications by appropriate
choices of terms for both the formation and dissolution processes, as
we show in Section~\ref{sec:example}.  Because it is based on
ERGMs, it will share in advances made on those models as well.  The
model is very useful for simulating realistic dynamic networks. This
is because of the sequential specification, the tractable parameters
and the relatively light computation burden.

As illustrated in Section~\ref{sec:example}, missing data on the
relational information can be dealt with in likelihood-based inference
using the approach of \citet{hangile10}. If the longitudinal data are
partially observed due to either a sample design or a missing data
process and is amenable to the model then their method is directly
applicable.

The assumption of within-step independence of formation and
dissolution is an important one, and its appropriateness depends on
the substantive setting and the basic nature of the process. Some
settings do not allow a separable formulation at all. For example an
affiliation network of players to teams in some sports, with a
realization observed during every game, imposes a hard constraint that
a player must belong to exactly one team at a time, and no team can
have more or fewer than a particular number of players, so the basic
unit of network change is teams trading players, rather than a player
joining or quitting a team. In settings that do allow simpler atomic
changes, separability may be a plausible approximation if the amount
of change between the discrete time steps is relatively small --- that
each time step represents a fairly small amount of time. As the length
of the time step increases, the separability approximation may become
less and less plausible. For example, a marriage network, even though
it has a hard constraint of each actor having at most one spouse at a
time, could be plausibly approximated in a separable framework (using, e.g., $\netsYF(\yat{-1})\equiv \{\y\in
2^\dysY:\y\supseteq\yat{-1}\land \forall_{i\in \actors}\abs{\y_i}\le 1 \}$) if each
discrete time step represented one month (since relatively few people
divorce and remarry in the same month) but not if it represented ten
years. More generally, the simpler the formation and dissolution processes
are within a time-step and the weaker the dependence between them,
the more plausible the assumption. (Of course, continuous-time Markov
models, to which these models asymptote, do not require an independence
assumption at all.)

As with the data used in Section~\ref{sec:example}, restriction on
the number of alters reportable is a common feature of network
surveys. Other examples of this censoring include the Add Health
friendship networks \citep{harris2003nls} and Sampson's monastery
data \citep{sampson1968npc}. To the extent that these are features of
the sampling design, they should be reflected in the likelihood. Per
Section~\ref{sec:sep-ergm}, a STERGM can be represented
as a TERGM \eqref{eq:dtergm}, which allows the sample space
$\netsY$ of $\Yat{}$ to be constrained to reflect this
design. Changing $\netsY$ only affects $\ceg(\curvpar)$ in
$l(\curvpar)$ --- the kernel of the model remains separable. This
situation is similar to that with censored data in survival analysis
where the likelihood is altered to reflect the censoring while the
model, and its interpretation, is unchanged.

Since assuming separability between formation and dissolution grants
significant benefits to interpretability, it would be useful to be
able to test if separability may be assumed in a given network
process. Some avenues for such tests include comparing goodness-of-fit
of a given model in modeling a transition $\y^0\to\y^2$ to its
modeling a transition $\y^0\to\y^1\to\y^2$ (with homogeneous
parameters). Or, if only one transition is available, a transition
$\y^0\to\y^1$ to a transition $\y^0\to\Y^\half\to\y^1$, with a latent
intermediate network $\Y^\half$. Development of such tests is beyond
the scope of this work and is subject for future research.

The STERGM framework allows a number of extensions to the model.
Over time, networks do not merely change ties:
actors enter and leave the network, and actors' own attributes
change. It is possible to incorporate the network size
adjustment developed by \citet{krivitsky2011ans} into these dynamic
models.
We have focused on longitudinal data.  It is possible to fit the
model based on egocentrically sampled data when the data includes
durational information on the relational ties \citep{krivitsky2008smd,krivitsky2009sms}.

\section*{Acknowledgments}
This work was supported by NIH awards R21 HD063000-01 and P30 AI27757,
NSF award MMS-0851555 and HSD07-021607, ONR award N00014-08-1-1015,
NICHD Grant 7R29HD034957, NIDA Grant 7R01DA012831, the University of
Washington Networks Project, and Portuguese Foundation for Science and
Technology Ci\^{e}ncia 2009 Program.  The authors would like to thank
the members of the University of Washington Network Modeling Group,
and especially Steven Goodreau for their helpful input.

\addcontentsline{toc}{section}{References}
\bibliographystyle{plainnat}
\bibliography{A_Separable_Model_for_Dynamic_Networks}

\appendix
\section{Separable TERGM Terms}\label{app:sdtergm-terms}
In this appendix we derive and discuss some fundamental model terms that can be used in a STERGM.
\gdef\thesection{A}
\subsection{Edge Counts}
\subsubsection{Formation}
Let $\genstatF(\yF,\yat{-1})=\abs{\yF}.$ This is equivalent to
$\genstat(\yat{},\yat{-1})=\abs{\yat{}\cup\yat{-1}}$. If
$\yatij{-1}=1$, $\yatij{}\lor\yatij{-1}=1$, so the state of $\yatij{}$
has no effect on $\genstat(\yat{},\yat{-1})$, but if $\yatij{-1}=0$,
$\yatij{}\lor\yatij{-1}=\yatij{}$, and the change in
$\genstat(\yat{},\yat{-1})$ is $1$.  This means that, in the absence
of other formation terms, $\curvparF$ represents the log-odds of a
given tie variable, that does not already have a tie, gaining a
tie. Then $\ilogit(\curvparF )$ is the expected fraction of tie
variables empty at time $t-1$ gaining a tie at time $t$. In the
presence of other terms, these log-odds become conditional
log-odds-ratios.

\subsubsection{Dissolution}
Let $\genstatD(\yD,\yat{-1})=\abs{\yD},$ or, equivalently,
$\genstat(\yat{},\yat{-1})=\abs{\yat{}\cap\yat{-1}}$. If
$\yatij{-1}=0$, $\yatij{}\land\yatij{-1}=0$, so the state of
$\yatij{}$ has no effect on $\genstat(\yat{},\yat{-1})$, but if
$\yatij{-1}=1$, $\yatij{}\land\yatij{-1}=\yatij{}$, and the change in
$\genstat(\yat{},\yat{-1})$ is $1$.  Then, in the absence of other
dissolution terms, $\curvparD $ represents the log-odds of a given tie
that exists at $t-1$ \emph{surviving} to $t$, and $\ilogit(\curvparD
)$ is the expected fraction of ties extant at time $t-1$ surviving to
time $t$. Depending on the problem, the interpretation of $-\curvparD
$ might be more useful: $\ilogit(-\curvparD )$ is the expected
fraction of extant ties being dissolved --- the hazard.

The formation phase can only affect non-tied pairs of actors, so if
the dissolution phase statistics have dyadic independence, the
formation process has no effect on duration distribution: in the
absence of other dissolution terms, the duration distribution of a tie
is geometric (with support $\NN$) with expected value
\citep[pp. 621--622]{casella2002si}
\[\left(\ilogit(-\curvparD )\right)^{-1}=1+\myexp{\curvparD}.\]

\subsection{Selective Mixing}
Selective mixing in the formation model can be represented by a vector
of statistics $\genstatF(\yF,\yat{-1})=|\yF_{P_{k_1},P_{k_2}}|$, with
notation described for \eqref{eq:churning-model}. However, in the
context of a STERGM, they have a direction.  
\subsubsection{Formation}
Let $\genstatF_{k_1,k_2}(\yF,\yat{-1})=|\yF_{P_{k_1},P_{k_2}}|$
(equivalently,
$\genstat_{k_1,k_2}(\yat{},\yat{-1})=|\yat{}_{P_{k_1},P_{k_2}}\cup\yat{-1}_{P_{k_1},P_{k_2}}|$).
The change in its value due to adding a tie $\pij$ (absent in
$\yat{-1}$) is $\I{i\in P_{k_1} \land j \in P_{k_2}}$, so
$\curvparF_{k_1,k_2}$ is the conditional log-odds-ratio due to the
effect of $i$ belonging to group $k_1$ and $j$ belonging to group
$k_2$ of a dyad $\pij$, that does not already have a tie, gaining a
tie. If the formation phase has no other terms, then the odds that
$\Yatij{}=1$ given that $\Yatij{-1}=0$ are
\[
\Oeg(\Yatij{}=1|\Yatij{-1}=0,i\in P_{k_1}\land j\in P_{k_2};\curvparF_{k_1,k_2})=\myexp{\curvparF _{k_1,k_2}}.
\]

\subsubsection{Dissolution}
Similar to the formation case, selective mixing can be represented by
a vector of statistics
$\genstatD(\yD,\yat{-1})=|\yD_{P_{k_1},P_{k_2}}|$. Then,
$g_{k_1,k_2}(\yat{},\yat{-1})=|\yat{}_{P_{k_1},P_{k_2}}\cap\yat{-1}_{P_{k_1},P_{k_2}}|$,
and $\curvparD _{k_1,k_2}$ is the conditional log-odds-ratio due to
the effect of $i$ belonging to group $k_1$ and $j$ belonging to group
$k_2$ of an extant tie $\pij$ being preserved until the next time
step.
\subsection{Degree Distribution}\label{sec:interp-degdist}
Unlike the first two examples, degree distribution statistics ---
counts of actors with a particular degree or range of degrees ---
introduce dyadic dependence into the model. As with many other such
terms, closed forms for many quantities of interest are not available,
and conditional log-odds are not as instructive, but the general
results for implicitly dynamic terms from Section~\ref{sec:sep-interp}
provide a useful heuristic, with the caveats discussed in that
section.

In practice, these terms are often used in conjunction with other
terms, so we only discuss their effect on the formation and
dissolution probabilities conditional on other terms --- their effect
over and above other terms, with those terms' coefficients held fixed.

\subsubsection{Formation}
Let $\y_i$ be the set of neighbors to whom $i$ has ties in $\y$. A
formation degree count term has the form
$\genstatF_k(\yF,\yat{-1})=\sum_{i\in\actors} \I{\abs{\yF_i}=d}$: the
number of actors $i$ in $\yF$ whose degree is $d$. The corresponding
TERGM statistic $\genstat_k(\yat{},\yat{-1})=\sum_{i\in\actors}
\I{\abs{\yat{}_i\cup\yat{-1}_i}=d}$.  We discuss the cases of $d=0$
and $d=1$, with the cases for $d > 1$ being similar to the $d=1$ case.
\begin{description*}
\item[$d=0$] By increasing the weight of those formation networks that
  have fewer isolates, a negative coefficient on this term increases
  the chances of a given actor gaining its first tie within a given
  time step. Conversely, a positive coefficient reduces the chances of
  an actor gaining its first tie. Because the term does not
  distinguish between different nonzero degrees, it mainly affects
  transitions from isolation to degree 1, not affecting further tie
  formation on that actor positively or negatively.
\item[$d=1$] Unlike the statistic for $d=0$, which can only be
  decreased by adding ties, the statistic for $d=1$ can be both
  increased and decreased (by making isolates into actors with degree
  1 and by making actors with degree 1 into actors with degree 2 and
  higher, respectively). Thus, the effect of this term is two-sided:
  with a positive coefficient, it both increases the chances of an
  actor gaining its first tie and reduces the chances of an actor
  gaining its second tie, while having relatively little effect on an
  actor with two ties gaining a third tie. A negative coefficient
  reduces the chances of an actor gaining its first tie, but if an
  actor already has one tie, it increases the chances that the actor
  gains a second tie.
\end{description*}

\subsubsection{Dissolution}
The analogous term in the dissolution model is
$\genstatD_k(\yD,\yat{-1})=\sum_{i\in\actors} \I{\abs{\yD_i}=d}$: same
as formation, but applied to $\yD$, and
$\genstat_k(\yat{},\yat{-1})=\sum_{i\in\actors}\I{\abs{\yat{}_i\cap\yat{-1}_i}=d}$.
\begin{description*}
\item[$d=0$] A negative coefficient on this term in the dissolution
  phase increases the weight of dissolution networks that have
  fewer isolates, and thus reduces the chances of a given actor losing
  its only tie, while a positive coefficient increases the
  chances of an actor losing its only tie. It may also have a modest
  effect on actors with more than one tie, since there is a positive
  probability of an actor losing more than one tie in the same time
  step.
\item[$d=1$] As in the case of formation, the effect of this term is
  two-sided: with a positive coefficient --- to preserve or create
  networks with more \enquote{monogamous} ties --- the chances of
  an actor losing its only tie decrease while the chances of an actor
  losing its second tie increase. (If an actor has 3 or more ties, the
  effect is weaker.)
  
  A negative coefficient on this term both increases the chances that
  an actor's last tie will be dissolved and reduces the chances that
  an actor with more than one tie has any ties dissolve.
\end{description*}

\subsection{\label{sec:interp-ergm}Other Standard Statistics}
Most statistics used in standard ERGM can be used in STERGM as
implicitly dynamic statistics.  For example, standard formation
statistics are
\begin{description*}
\item[Reciprocity:] $\sum_{\ijdysY,i<j} \yFij\yF_{j,i}$
\item[Transitive ties:] $\sum_{\ijdysY} \yFij\max_{k\in\actors}(\yF_{i,k}\yF_{k,j})$
\item[Cyclical ties:] $\sum_{\ijdysY} \yFij\max_{k\in\actors}(\yF_{k,i}\yF_{j,k})$
\item[Outdegree popularity (sqrt):] $\sum_{\ijdysY} \yFij\sqrt{\abs{\yF_{\cdot,j}}}$
\item[Edge covariate:] For a covariate $\x\in \RR^{n\times n}$, $\sum_{\ijdysY} \yFij\x\sij$
\end{description*}
The corresponding dissolution statistics have the same form, with $\yF$ replaced by $\yD$.

\end{document}